\documentclass[11pt]{article}
\usepackage{epsfig,amssymb,amsmath}
\usepackage{graphicx}

\begin{document}

\title{Complex dynamics near extinction in a predator-prey model with ratio dependence and Holling type III functional response}

\author{K.B. Blyuss\thanks{Corresponding author. Email: k.blyuss@sussex.ac.uk}, Y.N. Kyrychko \hspace{0.5cm} 
\\\\ Department of Mathematics, University of Sussex, Falmer,\\
Brighton, BN1 9QH, United Kingdom\\\\
\and O.B. Blyuss
\\\\ Wolfson Institute of Population Health, Queen Mary University of London,\\
London, EC1M 6BQ, United Kingdom}

\maketitle

\begin{abstract}

In this paper, we analyse a recently proposed predator–prey model with ratio dependence and Holling type III functional response, with particular emphasis on the dynamics close to extinction. By using Briot-Bouquet transformation we transform the model into a system, where the extinction steady state is represented by up to three distinct steady states, whose existence is determined by the values of appropriate Lambert W functions. We investigate how stability of extinction and coexistence steady states is affected by the rate of predation, predator fecundity, and the parameter characterising the strength of functional response. The results suggest that the extinction steady state can be stable for sufficiently high predation rate and for sufficiently small predator fecundity. Moreover, in certain parameter regimes, a stable extinction steady state can coexist with a stable prey-only equilibrium or with a stable coexistence equilibrium, and it is rather the initial conditions that determine whether prey and predator populations will be maintained at some steady level, or both of them will become extinct. Another possibility is for coexistence steady state to be unstable, in which case sustained periodic oscillations around it are observed. Numerical simulations are performed to illustrate the behaviour for all dynamical regimes, and in each case a corresponding phase plane of the transformed system is presented to show a correspondence with stable and unstable extinction steady state.

\end{abstract}

\section{Introduction}

A starting point for many ecological models analysing predator-prey interactions is provided by models of Gause-Kolmogorov type \cite{freedman,Gause,Kolm}
\[
\begin{array}{l}
\dot{u}=uf(u)-vg\left(u,v\right),\\\\
\dot{v}=bvg\left(u,v\right)-dv,
\end{array}
\]
where $u(t)$ and $v(t)$ are population densities or abundances of prey and predator, $d$ is the natural death rate of predators, and functions $f(u)$ and $g(u,v)$ describe, respectively, natural per capita growth rate of prey, and the trophic function or functional response of predators \cite{sol49}, which quantifies how efficiently predators are consuming prey, and how prey consumption increases predator reproduction. Function $f(\cdot)$ is often chosen in the form of either a constant (so-called Malthusian growth), or as a monotonically decreasing linear function $f(u)=r(1-u/K)$, which describes intra-specific competition of prey for available resources, and is known as the logistic growth of prey with a linear growth rate $r$ and the carrying capacity $K$. In the simplest case, the function $g(u,v)$ would only depend on the abundance of prey $g(u,v)=g(u)$, in which case it is known as prey-dependent functional response. Perhaps, the simplest example of prey-dependent functional response is when it is proportional to the number of prey $g(u)=au$, which yields the classical Lotka-Volterra model \cite{lotka,volt}. A more realistic representation of interactions between prey and predators is given by a functional response suggested by Holling \cite{hol1,hol2}, which accounts for the time predators spend searching for prey, as well as handling (i.e. chasing, killing and digesting) the prey. Holling proposed three types of functional response $g(u)$, which all satisfy $g(0)=0$ and approach some constant value for large $u$. The difference between different types of Holling functional response is in the behaviour for small prey numbers/densities. Type I response is linearly increasing for small prey densities, whereas for large prey numbers it saturates at some constant value; type II and type III are functions that are also saturating at high prey numbers, and are, respectively, concave and sigmoidal. Besides direct impact of predation, prey behaviour can also be affected by the fear of predation, which has been explored in a series of recent papers \cite{tiwari1,tiwari2,tiwari3,tiwari4,tiwari5}.

By the early 1990s, a number of ecological studies showed that when studying predator-prey interactions across a variety of scales and biological species, from insects to mammals \cite{rdexp1,rdexp2,rdexp3,rdexp4}, it may be more realistic to consider the functional response that depends on the ratio of prey to predators, which became known as ratio-dependent functional response \cite{ard_ginz,AGbook}. Such functional response can effectively model an observation that at higher predator densities, predators would have to compete for and share some of the prey. Formally, this would correspond to writing the functional response $g(u,v)$ in the form $g(u/v)$, which would then yield the following model
\[
\begin{array}{l}
\displaystyle{\dot{u}=uf(u)-vg\left(\frac{u}{v}\right),}\\\\
\displaystyle{\dot{v}=bvg\left(\frac{u}{v}\right)-dv.}
\end{array}
\]
A number of theoretical studies have studied the dynamics of models with ratio dependence \cite{RD1,RD2,RD3,RD4,RD5,RD6,RD7} and various forms of functional response that have usually taken the form of $g(z)=a$, $\displaystyle{g(z)=a_1 z/(1+a_2z)}$, and $\displaystyle{g(z)=a_1 z^n/(1+a_2z^n)}$ with $n>1$, where $z=u/v$, for Holling type I, II and III responses, respectively \cite{tyut,dawes}. Some additional ecological effects that have been studied in models with ratio dependence include time delays \cite{fan04,pal14,wang15}, as well as spatial dependence \cite{li08,apre10,guin14}. One important observation to make here is that due to the very nature of ratio-dependent functional response, the origin, i.e. the point $(u,v)=(0,0)$ that is characterised by the extinction of both species, may create certain mathematical challenges for analysis from the perspective that either the vector field itself is not defined at that point, or it is defined, but the linearisation is not, which hinders standard stability analysis. To overcome these challenges, one approach that is often used consists in rescaling time with the denominator of functional response, often in combination with Briot-Bouquet transformation to remove the singularity \cite{RD3,RD5,RD7}.

In this paper we are interested in Holling type III functional response with ratio dependence that has been observed in a number of experimental settings \cite{mor10,schenk05,sarn08,kratina08}. Rather than using a functional form $\displaystyle{g(z)=a_1 z^n/(1+a_2z^n)}$ given by a ratio of two polynomials, where $z=u/v$ and $n>1$, we consider an alternative form that can be written as follows,
\begin{equation}\label{TrR}
g(z)=ae^{-\alpha/z},
\end{equation}
which satisfies the conditions of $g(0)=0$, is monotonically increasing, and settles at a constant value as $z\to\infty$. Biological motivation for this form of functional response comes from the studies of plant-parasite interactions, in which plants serve as hosts for insect predators, and hence, current plant population represents a carrying capacity for insect population, in which case the interaction term has the form $g(u,v)=a(1-v/u)$ \cite{holt97,blyuss20}. Formally, trophic function (\ref{TrR}) is reminiscent of the Ivlev trophic function $g(z)=a\left(1-e^{-\alpha z}\right)$ \cite{ivlev} that represents Holling type II response, as well as of the Ricker model $N_{t+1}=N_t\exp[r(1-N_t)]$ for single-species populations \cite{ricker}. Making ratio dependence in (\ref{TrR}) explicit, we have the functional response $g(u,v)$ being given by
\begin{equation}
g(u,v)=ae^{-\alpha v/u}.
\end{equation}
A recent work has used a delayed version of this trophic function to analyse the dynamics of vector-plant interactions in the context of modelling plant mosaic disease \cite{fahad1}, while the role of stochastic effects in the time-delayed model has been studied in \cite{BKK21}. Of particular interest to us is the analysis of model behaviour in the neighbourhood of extinction $(u,v)=(0,0)$, and the effects this has on overall dynamics. While earlier numerical simulations have suggested that extinction is indeed possible in some parameter regimes \cite{BKK21}, the question of when exactly this happens, and whether extinction can coexist with other states has not been explored. 

The remainder of this paper is organized as follows. In the next section we identify different steady states of the model and show that the origin that corresponds to extinction of both species is a well-defined steady state. Applying Briot-Bouquet transformation, we will then obtain another version of our model, where the original extinction steady state is unfolded as an entire axis in the phase space. Whereas the origin of the modified model is always unstable, there are two more steady states that also correspond to 
the extinction steady state of the original model, and whose locations are given by a Lambert W function. We will derive analytical conditions for stability of these steady states and will illustrate regions of their feasibility and stability depending on model parameters. Model dynamics is further explored by numerically computing regions of feasibility and stability for different steady states and periodic orbits of the model depending on predation rate, predator growth rate, and the parameter characterising functional response. We also demonstrate phase plane of the modified model in each scenario, clearly indicating different steady states that correspond to extinction and coexistence, and complement these by numerical solutions of the original model, which can exhibit such distinct types of behaviour, as extinction of both species; survival of prey only, when predator growth rate is not sufficiently high; a regime of bi-stability between these two scenarios, where for the same values of parameters it is the initial values that determine whether both species or only predators will go extinct. Other possibilities include a stable coexistence steady state, where both prey and predators are maintained at some constant level, suggesting that predation is compensated or balanced by the prey growth, a periodic solution around the coexistence steady state, which is reminiscent of oscillations in the standard Lotka-Volterra model, and a regime of bi-stability between a stable coexistence and extinction. The paper concludes with a discussion of results.

\section{Methods}

As discussed in the Introduction, we consider a predator-prey model with logistic growth of prey and a Holling type III functional response with ratio dependence, which has the form
\begin{equation}\label{mod1}
\begin{array}{l}
\displaystyle{\dot{u}=ru\left(1-\frac{u}{K}\right)-ave^{-\alpha v/u},}\\\\
\displaystyle{\dot{v}=bve^{-\alpha v/u}-dv.}
\end{array}
\end{equation}
In order to simplify the model and to reduce the number of free parameters, we rescale the variables and parameters as follows,
\[
\begin{array}{l}
u=K\widehat{u},\quad v=K\widehat{v},\quad rt=\widehat{t},\\\\
\displaystyle{\frac{d}{r}=\widehat{d},\quad \frac{a}{r}=\widehat{a},\quad \frac{b}{r}=\widehat{b}.}
\end{array}
\]
The model (\ref{mod1}) can then be rewritten in the form
\begin{equation}\label{mod2}
\begin{array}{l}
\displaystyle{\dot{u}=u(1-u)-ave^{-\alpha v/u},}\\\\
\displaystyle{\dot{v}=bve^{-\alpha v/u}-dv,}
\end{array}
\end{equation}
where we have dropped hats for notational convenience. It is straightforward to show that this system is well-posed in that for arbitrary non-negative initial conditions, its solutions will remain non-negative and bounded for all $t\geq 0$. For any values of parameters, the system (\ref{mod2}) has a {\it prey-only steady state} $E=(1,0)$, which is stable for $b<d$ and unstable for $b>d$ \cite{BKK21}. If $b>d$ and $\alpha b-ad\ln(b/d)>0$, this system also has a {\it coexistence steady state} $E^*=(u^*,v^*)$, with
\[
\displaystyle{u^*=\frac{1}{\alpha b}\left[\alpha b-ad\ln\left(\frac{b}{d}\right)\right],
\quad v^*=\frac{1}{\alpha b}\ln\left(\frac{b}{d}\right)\left[\alpha b-ad\ln\left(\frac{b}{d}\right)\right].}
\]
This steady state is stable, provided the following condition is satisfied \cite{BKK21}
\[
-1+d\ln\left(\frac{b}{d}\right)\left[\frac{a}{\alpha b}\left(2-\ln\left(\frac{b}{d}\right)\right)-1\right]<0.
\]
In an earlier paper \cite{BKK21}, we focused on the analysis of coexistence steady state $E^*=(u^*,v^*)$ under effects of maturation time delay in predators, and also investigated the role of stochasticity.

\begin{figure}
\begin{center}
\includegraphics[width=0.6\textwidth]{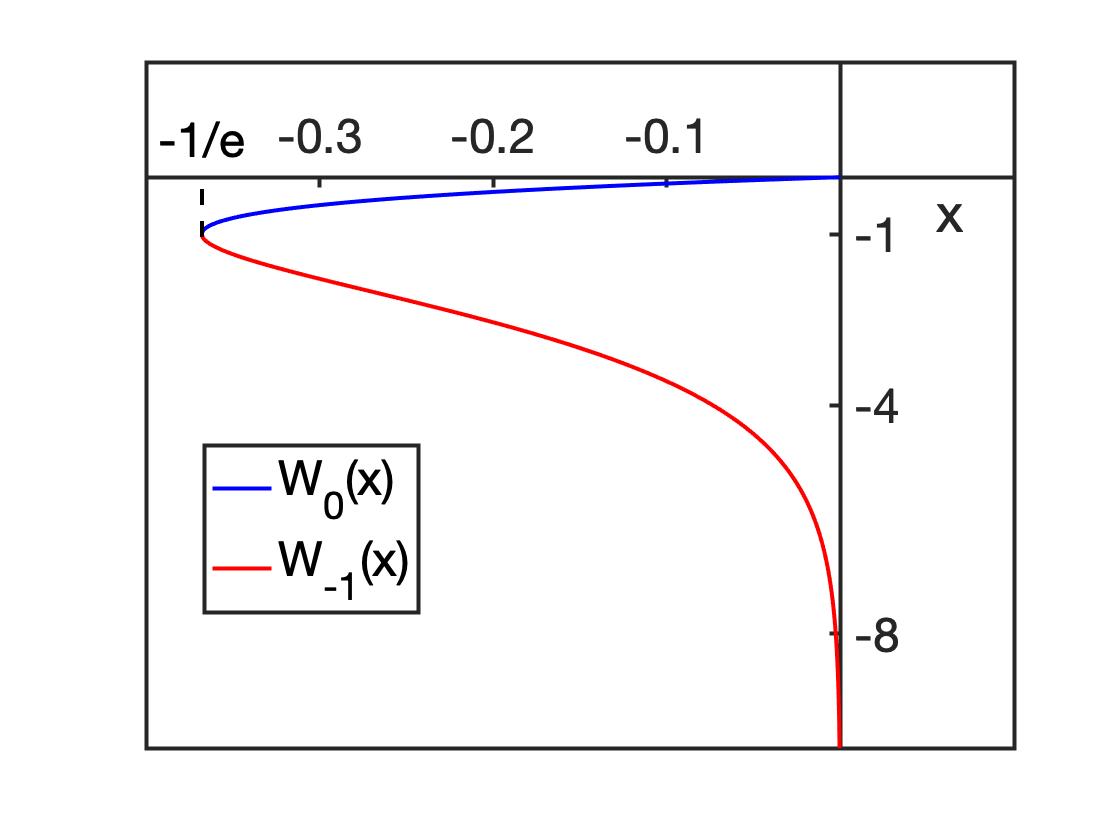}
\end{center}
\caption{Lambert functions $W_0(x)$ and $W_{-1}(x)$.}\label{lamb}
\end{figure}

In the neighbourhood of the point $(u,v)=(0,0)$, in the first quadrant the term $ve^{-\alpha v/u}$ is well-defined and positive, and in the limit $(u,v)\to (0,0)$, we have
\[
\displaystyle{0\leq \left | ve^{-\alpha v/u}\right |\leq |v|\xrightarrow[(u,v)\to (0,0)]{} 0,}
\]
which shows that $(0,0)$ is indeed a steady state of the system (\ref{mod2}), which biologically describes extinction of both species. Since the system (\ref{mod2}) is not differentiable at the point $(0,0)$, to explore its dynamics close to this steady state, we employ the Briot-Bouquet transformation \cite{zhang92} that in our case consists in introducing a new variable $z$, such that $v=zu$, and the system (\ref{mod2}) then transforms into
\begin{equation}
\begin{array}{l}\label{mod_sys}
\displaystyle{\dot{u}=u(1-u)-azue^{-\alpha z},}\\\\
\displaystyle{\dot{z}=z(b+az)e^{-\alpha z}-z(d+1-u).}
\end{array}
\end{equation}
The Briot-Bouquet transformation blows up the origin into an entire $z$-axis \cite{Dum06}. Hence, we should investigate equilibria of the system (\ref{mod_sys}) that lie on the $z$-axis.

For any parameter values, the transformed system (\ref{mod_sys}) has a steady state $E_0=(0,0)$. Besides this steady state, as long as the following condition is satisfied
\begin{equation}
\displaystyle{\frac{\alpha (d+1)e^{-b\alpha /a}}{a}\leq \frac{1}{e},}
\end{equation}
the system (\ref{mod_sys}) can have up to two more steady states $E_1=(0,z_1)$ and $E_2=(0,z_2)$,
where
\[
\displaystyle{z_1=-\frac{b}{a}-\frac{1}{\alpha}W_0\left[-\frac{\alpha (d+1)e^{-b\alpha /a}}{a}\right],\quad z_2=-\frac{b}{a}-\frac{1}{\alpha}W_{-1}\left[-\frac{\alpha (d+1)e^{-b\alpha /a}}{a}\right],}
\]
and $W_0(\cdot)$ and $W_{-1}(\cdot)$ are two branches of Lambert W function. Figure~\ref{lamb} illustrates the dependence of these two functions on their argument, and we note that the smallest possible value of $W_0(x)$ is attained at $x=-1/e$ and is equal to $W_0(-1/e)=-1$, which suggests that the steady state $E_1$ is only biologically feasible when $0<\alpha b/a<1$. Furthermore, since $W_{-1}(x)\leq W_0(x)<0$ for $x<0$, this implies that whenever both $E_1$ and $E_2$ exist, we have $z_1\leq z_2$.

Linearisation of the system (\ref{mod_sys}) near any steady state $\widehat{E}=(\widehat{u},\widehat{z})$ gives the Jacobian
\begin{equation}\label{Jac}
J=\left|
\begin{array}{cc}
1-2\widehat{u}-a\widehat{z}e^{-\alpha z} & -a\widehat{u}e^{-\alpha \widehat{z}}(1-\alpha \widehat{z})\\
\widehat{z} & (b+a\widehat{z})(1-\alpha \widehat{z})e^{-\alpha \widehat{z}}+aze^{-\alpha \widehat{z}}-(d+1-\widehat{u})
\end{array}
\right|.
\end{equation}
At $E_0$, the two eigenvalues are $1$ and $b-d+1$, suggesting that this steady state is always unstable and is either a saddle, or an unstable node. At the steady states $E_1$ and $E_2$, we have the Routh-Hurwitz conditions for stability in the form
\[
1<a\widehat{z}e^{-\alpha \widehat{z}}<\alpha(d+1)\widehat{z}.
\]
\newpage

\begin{figure}[h!]
\includegraphics[width=\textwidth] {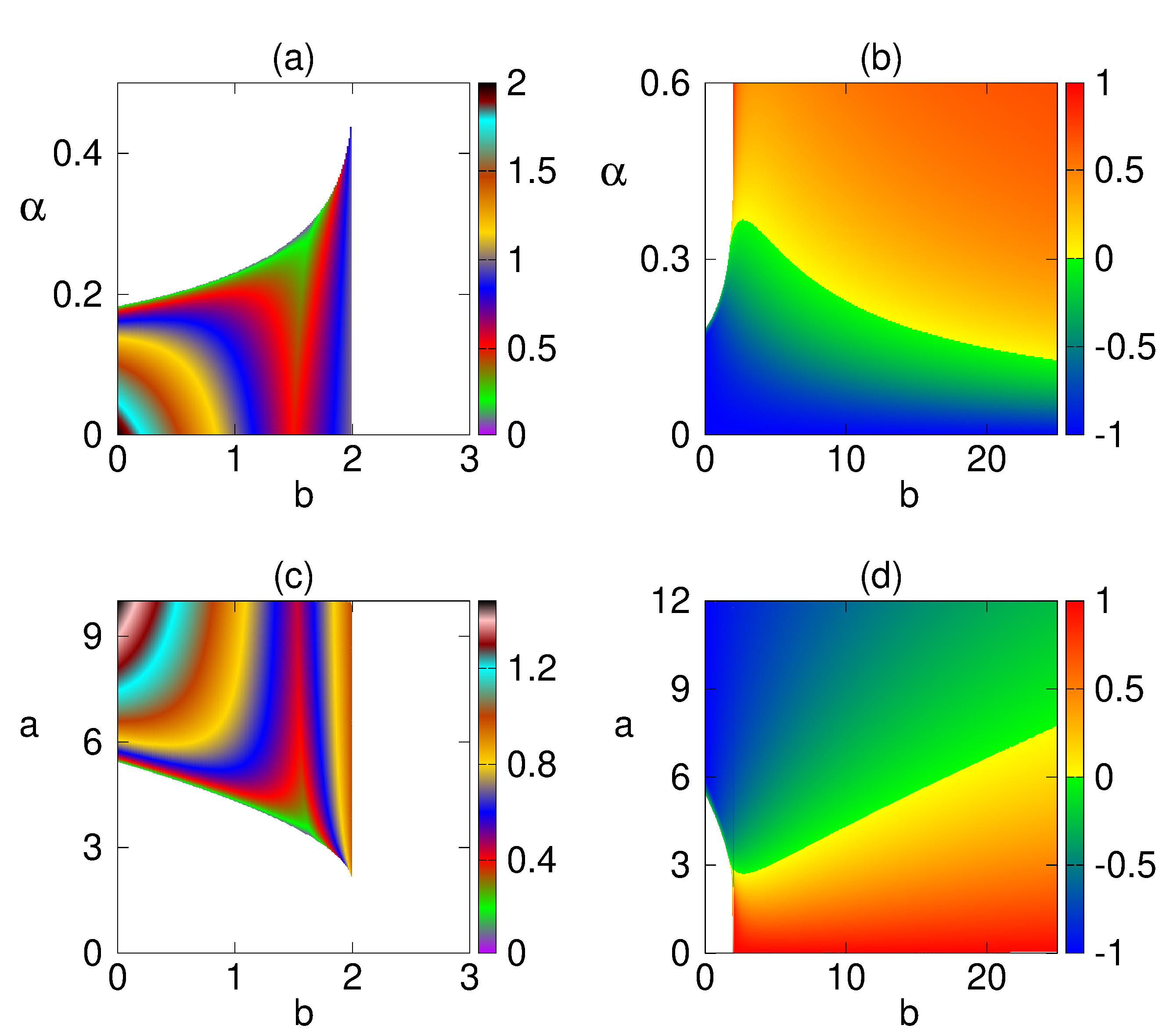}
\vspace{-0.7cm}
\caption{Regions of feasibility and stability of steady states $E_1$ (a,c) and $E_2$ (b,d). Steady states are feasible in a coloured region, and biologically infeasible in the white region. Colour code denotes the largest real part of characteristic eigenvalues. Parameter values are $a=d=1$ (a,b), $\alpha=d=1$ (c,d).}\label{stab}
\end{figure}

\section{Results}

In order to illustrate the effects of different parameters on feasibility and stability of the two steady states $E_1$ and $E_2$ that correspond to extinction in the original model, we plot in Fig.~\ref{stab} maximum real part of the characteristic eigenvalues as determined by the Jacobian (\ref{Jac}). We observe that the steady state $E_1$ only exists for sufficiently small levels of predator growth rate $b$, and either low enough predation rate $a$, or sufficiently small rate $\alpha$. For any parameter combination, where this steady state is biologically feasible, it is unstable. In contrast, starting with rather small values of $b$, the steady state $E_2$ is biologically feasible for an entire range of values of $a$ or $\alpha$ and is stable for smaller values of $a$, or for larger values of $\alpha$, and the region of stability shrinks with increasing predator growth rate $b$. 

\newpage

\begin{figure}[h!]
\includegraphics[width=\textwidth] {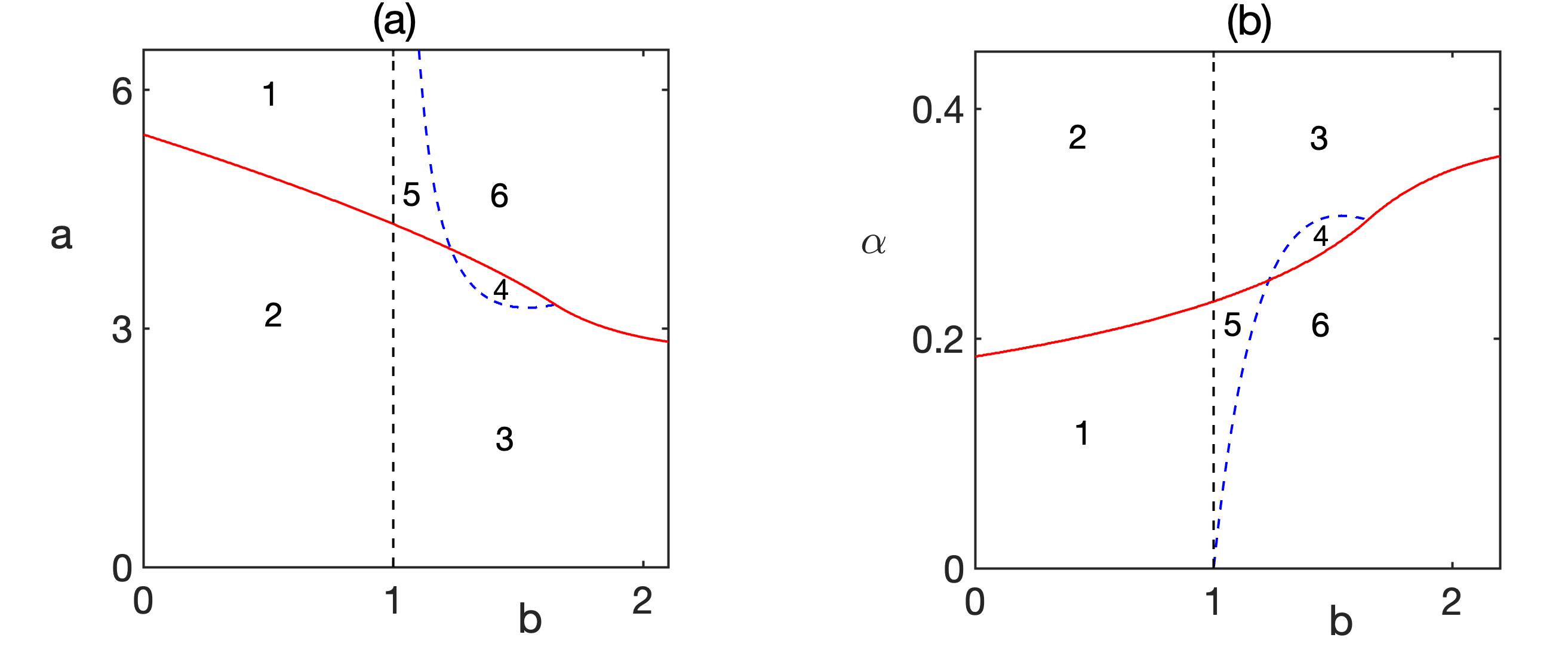}
\caption{Regions of feasibility and stability of different steady states of the model (\ref{mod2}). The prey-only steady state $E=(1,0)$ is stable to the left of the dashed black line at $b=1$, and unstable to the right of that line. The extinction steady state $E_0$ is stable above red line in plot (a) and below red line in plot (b). In region 1, there is bi-stability between a stable steady state $E$ and the extinction steady state $E_0$; in region 2 only the steady state $E$ is stable; in region 3, only the coexistence steady state $E^*$ is stable; in region 4, the coexistence steady state $E^*$ is unstable, and there is periodic orbit around this state; in region 5, there is bi-stability between the extinction steady state $E_0$ and the coexistence steady state $E^*$; in region 6, only extinction steady state $E_0$ is stable. Parameter values are $\alpha=d=1$ (a), $a=d=1$ (b).}\label{pplane}
\end{figure}

Under the inverse Brio-Bouquet transformation, we can interpret regions of stability of the steady state $E_2$ of the transformed system (\ref{mod_sys}) as parameter regions, where the extinction steady state $E_0=(0,0)$ of the original system (\ref{mod2}) is stable. Combining these results with those on stability of the prey-only equilibrium $E$ and the coexistence equilibrium $E^*$, we now illustrate in Fig.~\ref{pplane} an entire range of dynamical scenarios that can be exhibited in the model. 

\begin{figure}[h!]
\vspace{-0.4cm}
\includegraphics[width=\textwidth] {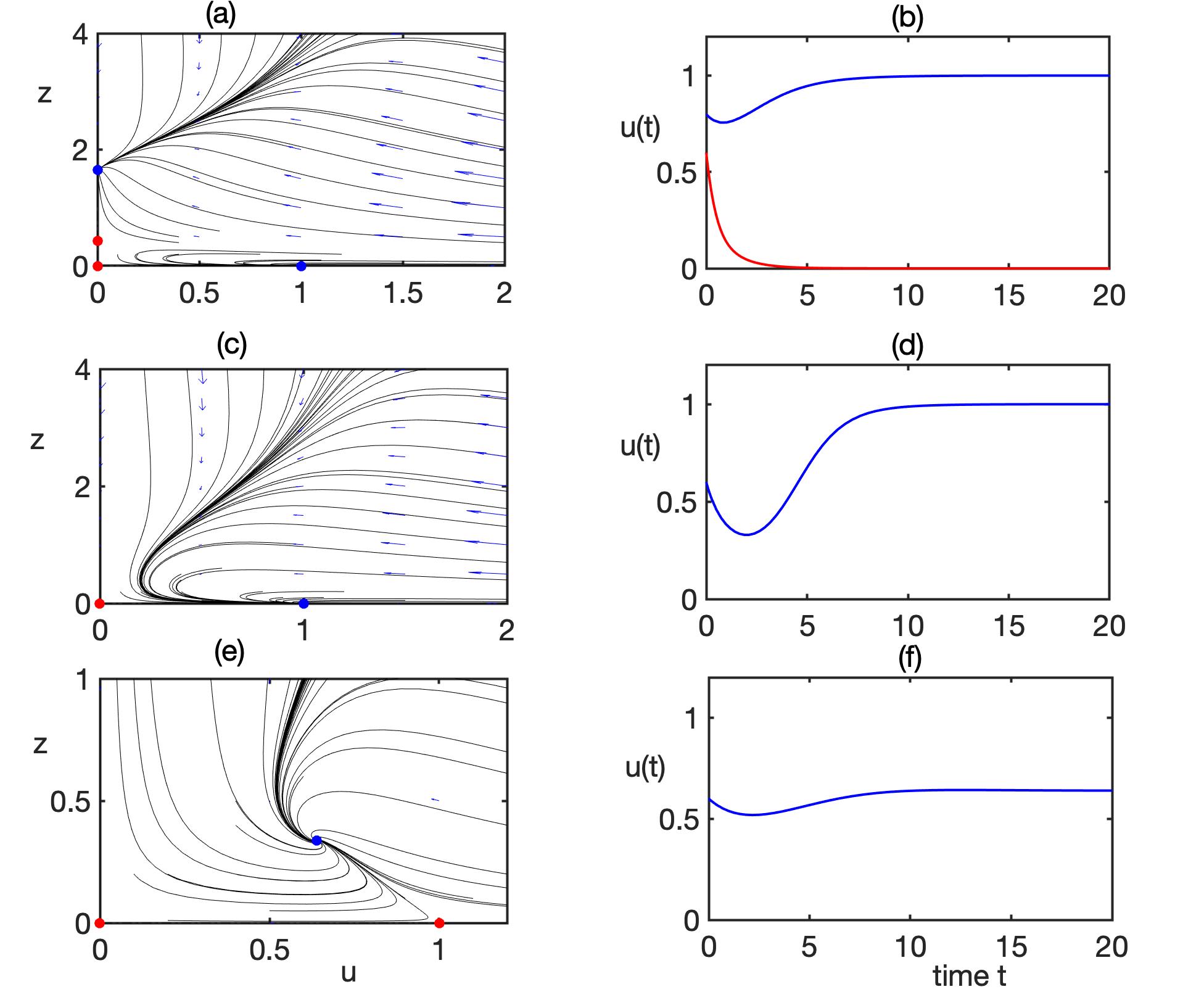}
\vspace{-0.7cm}
\caption{Phase planes (left column) of the transformed system (\ref{mod_sys}) and the corresponding numerical solution (right column) of the rescaled original system (\ref{mod2}), illustrating different dynamical regimes listed in Fig.~\ref{pplane}(a). Blue/red dots in phase plane plots indicate stable/unstable steady states. Parameter values are $\alpha=d=1$. (a)-(b) Case 1, $a=6$, $b=0.5$, bi-stability between prey-only steady state $E$ and the extinction steady state $E_0$. (c)-(d) Case 2, $a=3$, $b=0.5$, prey-only steady state $E$ is the only stable state. (e)-(f) Case 3, $a=1.5$, $b=1.4$, only the coexistence steady state $E^*$ is stable.}\label{cases13}
\end{figure}

For values of $b$ smaller than $1$, which biologically describes a scenario, where the rate of growth in the population of predators is too small to be able to maintain their population as based on the available prey, the coexistence steady state $E^*$ is not biologically feasible, while the prey-only steady state $E$ is stable. In this case, the system either goes to this prey-only steady state as the only stable equilibrium for parameter values in region 1, or it exhibits a bi-stability between this steady state, and a stable extinction steady state $E_0$ in region 2. This is shown in detail in Fig.~\ref{cases13}, where, for convenience, for each parameter combination, in the right column we plot numerical solution of the model (\ref{mod2}), and in the left column we plot the corresponding phase plane of the transformed system (\ref{mod_sys}) to demonstrate how changes in stability of the extinction steady state $E_{0}$ affect global dynamics. Since in region 1 there is bi-stability between two steady states, both of which are characterised by the absence of predators, this suggests that it is rather the initial conditions of the system that determine whether prey will survive (though, initially, prey population will decline due to predation), or will also be driven to extinction. Interestingly, even though in this parameter regime, predator population goes to zero, this does not automatically guarantee the survival of the prey, which is affected not only by the values of parameters, but also by the initial presence of sufficiently many prey to maintain its stable population.

\newpage

\begin{figure}
\includegraphics[width=\textwidth] {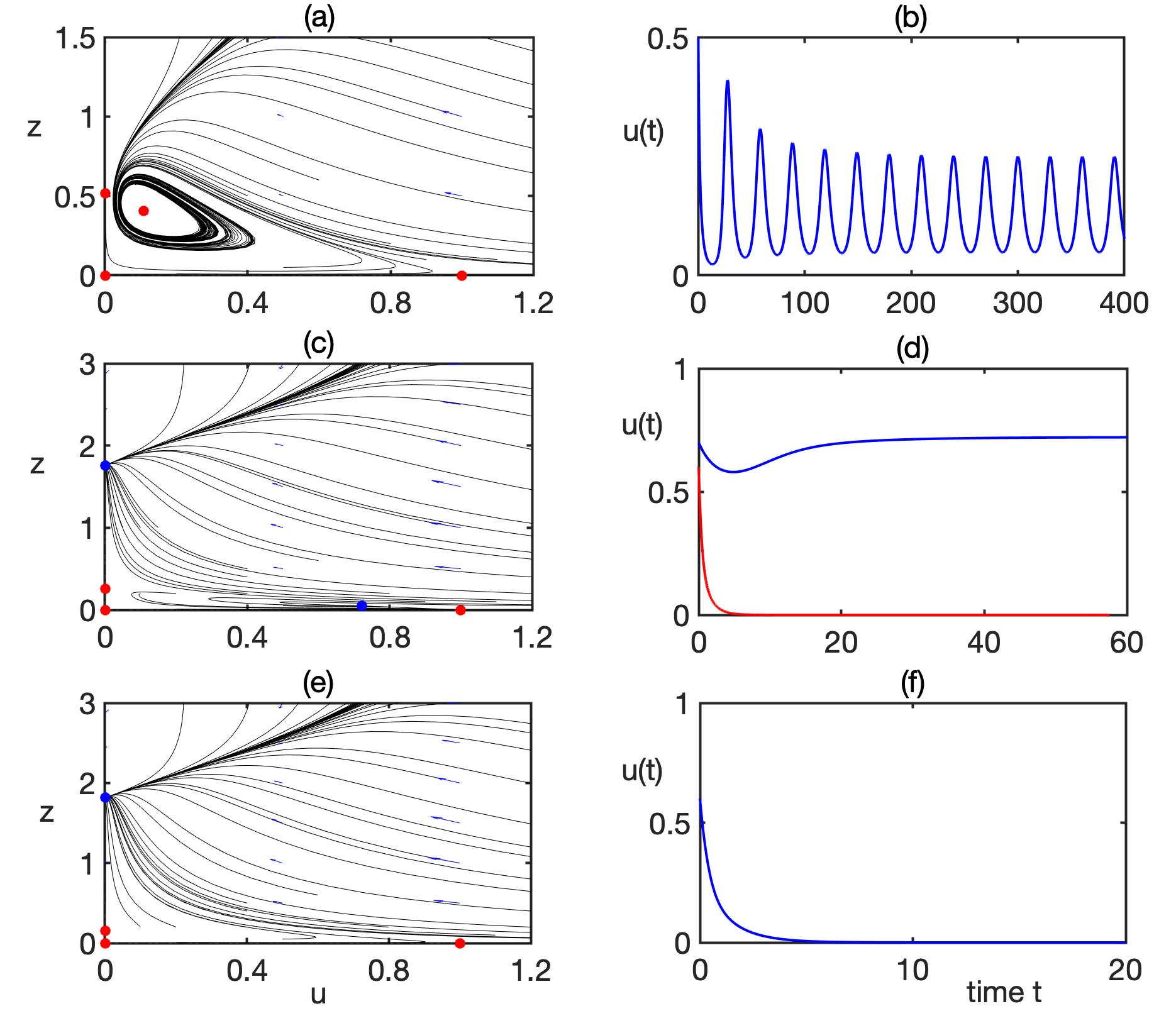}
\vspace{-0.7cm}
\caption{Phase planes (left column) of the transformed system (\ref{mod_sys}) and the corresponding numerical solution (right column) of the rescaled original system (\ref{mod2}), illustrating different dynamical regimes listed in Fig.~\ref{pplane}(a). Blue/red dots in phase plane plots indicate stable/unstable steady states. Parameter values are $\alpha=d=1$. (a)-(b) Case 4, $a=3.3$, $b=1.5$, periodic solution around the unstable coexistence steady state $E^*$. (c)-(d) Case 5, $a=6$, $b=1.05$, bi-stability between the extinction steady state $E_0$ and the coexistence steady state $E^*$. (e)-(f) Case 6, $a=6$, $b=1.4$, only the extinction steady state $E_0$ is stable.}\label{cases46}
\end{figure}

As the value of $b$ exceeds 1, the prey-only equilibrium $E$ becomes unstable, while the coexistence steady state $E^*$ becomes feasible and stable, as shown in Fig.~\ref{cases13}(e)-(f) (region 3). As the value of predation rate $a$ increases, as long it stays below the boundary of stability of the extinction steady state $E_0$, the coexistence steady state $E^*$ loses its stability through a supercritical Hopf bifurcation, giving rise to stable periodic solutions around this steady state, as illustrated in Fig.~\ref{cases46}(a)-(b) (region 4). Once the stability boundary of the extinction steady state is crossed, for smaller values of $b$ and sufficiently high values of predation rate $a$, stable coexistence steady state $E^*$ co-exists with a stable extinction steady state $E_0$ shown in Fig.~\ref{cases46}(c)-(d) (region 5). Biologically, this means that for the same parameter values, whether or not the system evolves towards coexistence of prey and predators, or they drive each other to extinction, is again determined by the choice of initial conditions. In region 6, which corresponds to high values of both predation rate and the growth rate of predators, the only stable state of the system is that of extinction, as illustrated in Fig.~\ref{cases46}(e)-(f).

\section{Discussion}\label{disc}

In this article, we have studied the dynamics of a predator-prey model with Holling type III functional response and ratio dependence. Earlier numerical results for the same model with maturation delay suggested the possibility of extinction of both prey and predator populations that was, however, not investigated at that time \cite{BKK21}. Whereas extinction steady state is a feasible steady state of the model for any values of parameters, due to ratio dependence in the functional response, the model could not be linearised in the neighbourhood of extinction steady state in the standard manner. To overcome this difficulty, we have transformed the system using a Briot-Bouquet transformation, which resulted in the unfolding of the extinction steady state into three distinct steady states, whose stability could now be studied. Two of those steady states are unstable for any values of parameters, while the third is stable for sufficiently large predation rate to have an effect on prey who are reproducing logistically, and for sufficiently small values of the parameter $\alpha$ characterising functional response. In both cases, we note that the region of stability of extinction steady state is shrinking for higher rates of predator fecundity, which can be explained by the fact that if for the same predation rate predators are better able to reproduce, this increases the possibility of a stable coexistence, where both populations are maintained at some steady level. Regions of feasibility of two steady states of the transformed model that correspond to extinction in the original model are determined explicitly in terms of system parameters, since they are given in the form of principal and $n=-1$ branches of the, in general, complex-valued Lambert W function.

By combining the results on stability of extinction steady state with conditions for stability of prey-only and coexistence equilibria, we are able to obtain a full picture of system dynamics in different parameter regions, identifying situations where only one of those steady states is stable, as well as cases of bi-stability, where extinction coexists with either a prey-only, or coexistence equilibria. In the parameter region, where the system exhibits sustained periodic oscillations around an unstable coexistence equilibrium, resulting from a supercritical Hopf bifurcation of this steady state, the extinction steady state is unstable, hence, no bi-stability is observed. While coexistence equilibrium is only biologically feasible, when the prey-only steady state is unstable, in contrast, the extinction steady state exists for all values of parameters and can coexist with either of those steady states. In the case of coexistence of steady states, for the same parameter values, depending on the initial conditions, the system will approach one of the two coexisting steady states. This observation is important from the perspective of analysing real ecological data, as it provides clues about the possibility of different types of dynamics for the same or very similar values of parameters, some of which may be difficult to accurately measure. Earlier papers that focused largely on ratio dependence in models with Holling type II functional response \cite{RD3,RD5,RD7} primarily showed either extinction of both species, or their stable coexistence. In contrast, our model with Holling type III response exhibits  significant parameter regions, where prey-only equilibrium is stable either by itself, or in combination with extinction steady state. On the other hand, in the case when coexistence steady state is unstable, and there is a periodic solution around it, it is the only attractor in the model.

There are several directions, in which the work presented in this paper could be extended. In an earlier paper \cite{BKK21}, we studied the effects of stochasticity and maturation delays for the same model, and some of the simulations there also indicated the possibility of observing extinction in the time-delayed model. Unfortunately, the standard Briot-Bouquet transformation, as used in this paper, would not work for the analysis of stability of extinction steady state in the time-delayed model, because even after the transformation, the singularity at the point $(0,0)$ would remain, and some alternative approach for studying stability of extinction equilibrium in a model with time delay needs to be developed. Another feature of many ecological models is the so-called stochastic extinction, where for sufficiently small population densities, species can go extinct, even though deterministically they could recover to some sustained levels at a later stage. It would be interesting and important to investigate how (in)stability of extinction steady state is affected by stochasticity, and how it impacts the regimes of bi-stability between extinction and other equilibria. This could be achieved through numerical exploration of basins of attraction of different steady states in the stochastic model in a manner similar to how it was done by Fatehi et al. \cite{fatehi1,fatehi2} in the context of modelling autoimmunity arising from immune response to a viral infection.

\section*{Conflict of Interest Statement}

The authors declare that the research was conducted in the absence of any commercial or financial relationships that could be construed as a potential conflict of interest.

\section*{Author Contributions}

All authors listed have made a substantial, direct, and intellectual contribution to the work and approved it for publication.

\section*{Funding}

OB has been supported by Cancer Research UK (EDDCPJT\textbackslash 100022).

\bibliographystyle{ieeetr}

\bibliography{test}

\end{document}